\documentclass[a4paper,fleqn]{cas-dc}
\usepackage{amssymb}
\usepackage{hyperref}
\usepackage{graphicx}
\usepackage[caption=false]{subfig}
\usepackage{lipsum}
\usepackage{lscape}
\usepackage{amsmath}
\usepackage{multirow}
\usepackage[figuresright]{rotating}
\usepackage{lineno}
\usepackage{algorithm} 
\usepackage{algpseudocode} 
\usepackage{lineno}
\usepackage{nomencl}
\makenomenclature
\usepackage{calc}
\usepackage[T1]{fontenc}
\usepackage{relsize}

\usepackage[sort,comma,authoryear,round]{natbib}

\setcitestyle{authoryear,open={(},close={)},citesep={;}} 

\hypersetup{
  colorlinks,
  citecolor=Violet,
  linkcolor=Red,
  urlcolor=Blue}

\newcolumntype{P}[1]{>{\centering\arraybackslash}p{#1}}

\def\tsc#1{\csdef{#1}{\textsc{\lowercase{#1}}\xspace}}
\tsc{WGM}
\tsc{QE}
\tsc{EP}
\tsc{PMS}
\tsc{BEC}
\tsc{DE}

\begin{document}
\let\WriteBookmarks\relax
\def\floatpagepagefraction{1}
\def\textpagefraction{.001}
\shorttitle{Deep Reinforcement Learning  for Crop Production Management}
\shortauthors{Balderas et~al.}

\title [mode = title]{A Comparative Study of Deep Reinforcement Learning for Crop Production Management}

\author[1]{Joseph Balderas}\ead{joseph.balderas@uta.edu}
\author[2]{Dong Chen}\ead{dc2528@msstate.edu}
\author[3]{Yanbo Huang*}\ead{yanbo.huang@usda.gov}
\author[1]{Li Wang}\ead{li.wang@uta.edu}
\author[1]{Ren-Cang Li}\ead{rcli@uta.edu}

\address[1]{Department of Mathematics, University of Texas at Arlington, Arlington, TX 76019, USA}
\address[2]{Department of Agricultural \& Biological Engineering, Mississippi State University, Mississippi State, MS 39762, USA}
\address[3]{USDA-ARS Genetics and Sustainable Agriculture Research Unit, Mississippi State, MS 39762, USA}
\address{* Yanbo Huang (\textit{Yanbo.Huang@usda.gov}) is the corresponding author.}

\begin{abstract}
Crop production management is essential for optimizing yield and minimizing a field's environmental impact to crop fields, yet it remains challenging due to the complex and stochastic processes involved. Recently, researchers have turned to machine learning to address these complexities. Specifically, reinforcement learning (RL), a cutting-edge approach designed to learn optimal decision-making strategies through trial and error in dynamic environments, has emerged as a promising tool for developing adaptive crop management policies. RL models aim to optimize long-term rewards by continuously interacting with the environment, making them well-suited for tackling the uncertainties and variability inherent in crop management. Studies have shown that RL can generate crop management policies that compete with, and even outperform, expert-designed policies within simulation-based crop models. In the gym-DSSAT crop model environment, one of the most widely used simulators for crop management, proximal policy optimization (PPO) and deep Q-networks (DQN) have shown promising results. However, these methods have not yet been systematically evaluated under identical conditions. In this study, we evaluated PPO and DQN against static baseline policies across three different RL tasks, fertilization, irrigation, and mixed management, provided by the gym-DSSAT environment. To ensure a fair comparison, we used consistent default parameters, identical reward functions, and the same environment settings. Our results indicate that PPO outperforms DQN in fertilization and irrigation tasks, while DQN excels in the mixed management task. This comparative analysis provides critical insights into the strengths and limitations of each approach, advancing the development of more effective RL-based crop management strategies.
\end{abstract}

\maketitle

\section{Introduction}
\label{sec:intro}
Crop production management is the process of taking logical actions on a crop field in order to achieve crop production goals. Unfortunately, it can be difficult for farmers to choose optimal practices due to complex physical, chemical, and biological phenomena which take place on a crop field and random external factors, such as weather \citep{husson2021soil}. With the advent of machine learning and its success in the field of agricultural science \citep{liakos2018machine}, researchers have begun to study how data-intensive algorithms can be used to learn optimal policies for crop management \citep{ gautron2022reinforcement}. In particular, methods such as reinforcement learning (RL) have been applied to crop management problems and have produced competitive results against existing expert policies on crop simulators \citep{gautron2022gym, tao2022optimizing}. RL algorithms work by training an agent on a trial-and-error basis in an environment with the objective of maximizing cumulative returns; after training, the algorithms output static policies that can be deployed for real-world applications. 

Recent research has focused on the development of crop field RL environment software and the exploration of how different RL algorithms perform when applied to these virtual environments. One of the first crop RL environments introduced was CropGym, an OpenAI gym environment developed by \citet{overweg2021cropgym} for the problem of Nitrogen fertilization management. Proximal policy optimization (PPO) was used in this study to successfully learn fertilization policies which lowered negative environmental impact. \citet{gautron2022gym} then developed gym-DSSAT, a flexible gym environment that utilized the popular crop simulator Decision Support System for Agrotechnology Transfer (DSSAT) to support fertilization and irrigation problems. PPO was shown to outperform expert baseline policies in the gym-DSSAT environment. To address long-term policy learning, \citet{turchetta2022learning} introduced CyclesGym, a crop environment based on multi-year, multi-crop CGM Cyles. This RL environment was also tested using PPO. \citet{tao2022optimizing} implemented deep Q-network (DQN) on the gym-DSSAT environment to attain improved RL agent performance. They also implemented imitation learning (IL) for the scenario where only a few state variables are available. \citet{wu2022optimizing} continued this study by testing DQN against soft actor-critic (SAC) and other baseline policies on gym-DSSAT for the Nitrogen management problem. Finally, \citet{wu2024new} combined a language model (LM) with DQN to obtain a more optimal policy on the gym-DSSAT environment. This was achieved by using the LM to convert state variables into more informative language. While these studies demonstrated promising results, they primarily focused on evaluating individual RL algorithms, without systematically comparing different methods under consistent settings.

In this study, we conduct a comprehensive comparison between the PPO and DQN algorithms, focusing exclusively on RL within the gym-DSSAT environment. To ensure a fair comparison, we evaluate both algorithms using the reward functions and baseline methods from \citet{gautron2022gym} for the fertilization and irrigation tasks. For the mixed problem, which considers fertilization and irrigation simultaneously, we apply the economic profit reward function from \citet{tao2022optimizing}. Additionally, both models are trained with random weather conditions, addressing a recommendation from previous DQN studies \citep{tao2022optimizing, wu2022optimizing, wu2024new}, which used non-random weather but suggested randomization for future research. This comparison aims to provide valuable insights into the relative strengths of these algorithms, helping to inform the selection of appropriate RL methods for real-world crop management applications.

The remainder of this paper is structured as follows. First, we introduce the fundamental concepts of RL and provide a detailed overview of the competing algorithms, DQN and PPO. Next, we describe the gym-DSSAT environment, including its default settings and the three RL tasks it supports: fertilization, irrigation, and mixed management. Finally, we present and analyze the results of our experiments, followed by a discussion of key findings and suggestions for future research directions.

\section{Preliminaries of Reinforcement Learning}
\label{sec:RLbasics}
In this section, we review some preliminaries of RL, focusing on Markov Decision Process, and key RL algorithms.

\subsection{Markov Decision Process} 
An RL environment can be modeled as a Markov decision process (MDP) \citep{sutton2018reinforcement}. An MDP is defined by the tuple $\mathcal{M}=\langle S, A, \textbf{p}, \textbf{r} \rangle$, where $S$ is the set of environment states, $A$ is the set of actions, $\textbf{p}$ is the transition function, which gives the probability $\textbf{p}(s'|s, a)$ of transitioning to state $s' \in S$ given action $a \in A$ is taken in state $s \in S$, and $\textbf{r}$ is the reward function, which provides the expected reward $\textbf{r}(s, a, s') = \mathbb{E}[r|s, a, s']$ when action $a$ is performed in state $s$ and results in state $s'$ \citep{puterman2014markov}. At each time step $t \in {0,1,2,\dots, T-1}$, where $T$ can be infinite, an agent interacts with the environment by observing the current state $s_t$, taking an action $a_t$, receiving a reward $r_t$, and observing the next state $s_{t+1}$. A trajectory is the full sequence of states, actions, and rewards, denoted by $\tau = (s_0, a_0, r_0, s_1, a_1, r_1, \dots, s_T)$. The set of final states, $S_f \subset S$, consists of states that terminate an episode when reached.

\begin{figure}[ht]
\centering
\includegraphics[width=0.45\textwidth]{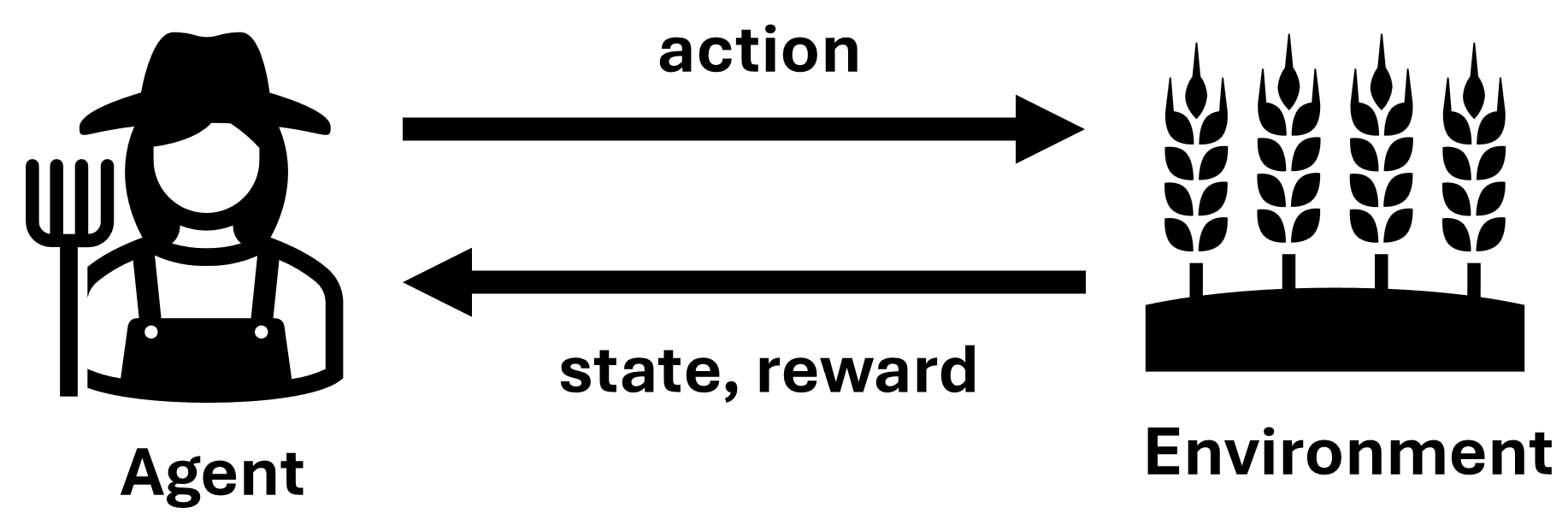}
\caption{In the RL process, an agent makes an action in an environment, and the environment in turn produces a new state and a reward which informs the agent of its current performance. The goal of the agent is to use the environment feedback to maximize its cumulative rewards. This loop repeats up to a specified number of iterations or until a terminal state is reached \citep{gautron2022reinforcement}.}
\label{fig:figure0}
\end{figure}

An MDP, together with an objective function $J$ that the agent seeks to optimize, is referred to as a Markov decision problem \citep{gautron2022reinforcement}. The objective function $J$ is defined as the expected discounted return:
\begin{equation} \label{eq:1}
    J(\pi) = \mathbb{E}_{\tau \sim p_\pi}\left[ \sum_{t=0}^{T-1} \gamma^t r_{t}\right],
\end{equation}
where $\pi: S \rightarrow \mathcal{P}(A)$ is a policy mapping each state $s$ to a probability distribution $\pi(a|s)$ over actions in $A$, $p_\pi$ is the trajectory distribution under policy $\pi$, and $\gamma \in (0,1]$ is a discount factor that reduces the influence of future rewards \citep{levine2020offline}. The goal of RL is to learn an optimal policy $\pi^*$ that maximizes $J(\pi)$. The policy is often modeled using a neural network with parameters $\theta$.

In RL, the value function $V^{\pi}: S \rightarrow \mathbb{R}$ assigns a value to each state $s$, representing the expected future rewards when starting from state $s$ and following policy $\pi$. The value function $V^{\pi}$ is formally defined as
\begin{equation}\label{eq:3}
    V^{\pi}(s) = \mathbb{E}_{\tau \sim p_\pi}\left[ \sum_{t=0}^{T-1} \gamma^{t} r_{t} \middle| s_0=s \right].
\end{equation}
Similarly, the state-action value function, or Q-function $Q^{\pi}: S \times A \rightarrow \mathbb{R}$, assigns a value to each state-action pair $(s, a)$, describing the expected future rewards when taking action $a$ in state $s$ and following policy $\pi$. It is defined as
\begin{equation}\label{eq:4}
    Q^{\pi}(s,a) = \mathbb{E}_{\tau \sim p_\pi}\left[ \sum_{t=0}^{T-1} \gamma^{t} r_{t} \middle| s_0=s, a_0=a \right].
\end{equation}

An optimal policy $\pi^*$ can be derived by first learning the optimal Q-function $Q^*$ and then choosing the action that maximizes this function:
\begin{equation}\label{eq:5}
\pi^*(a|s)=
    \begin{cases}
        1 & \text{if } a = \underset{a'\in A}{\text{argmax\hspace{0.1cm}}}Q^*(s,a'),\\
        0 & \text{otherwise}.
    \end{cases}
\end{equation}
An RL algorithm that combines policy gradient estimation with Q-learning is referred to as an \emph{actor-critic method}. In the following section, we will explore how DQN operates as a Q-learning method, while PPO serves as an actor-critic method.

\subsection{DQN algorithm}
Deep Q-Network (DQN) \citep{mnih2015human} is a breakthrough RL algorithm designed to approximate the optimal Q-function in environments with a discrete action space using a neural network, or Q-network. The objective is to approximate the optimal action-value function $Q^*(s, a)$, which is defined as
\begin{equation}\label{eq:6}
    Q^*(s,a) = \underset{\pi}{\text{max }} \mathbb{E}_{\tau \sim p_\pi}\left[ \sum_{t=0}^{T-1} \gamma^{t} r_{t} \middle| s_0=s, a_0=a \right],
\end{equation}
Once the optimal Q-function is learned, the agent follows a greedy policy by selecting the action that maximizes $Q^*(s, a)$ for each state, as shown in (\ref{eq:5}). The key equation for learning $Q^*$ is the Bellman equation \citep{sutton2018reinforcement}:
\begin{equation}\label{eq:7}
    Q^*(s,a) = \mathbb{E}_{s' \sim \textbf{p}}\left[r + \gamma \underset{a'}{ \text{ max }}Q^*(s',a') \middle| s,a \right],
\end{equation}
where $s'$ is the next state.

At each training step, the Q-network is updated by minimizing the mean-squared error between the predicted Q-values and the target values derived from the Bellman equation \citep{mnih2015human}. The loss function used to update the Q-network at iteration $i$ is defined as
\begin{equation}
    L_i(\theta_i)=\mathbb{E}_{s,a,r,s'}\left[\left(y_i-Q(s,a;\theta_i)\right)^2 \right],
\end{equation}
where $y_i = r + \gamma \underset{a'}{ \text{ max }}Q(s', a';\theta_i^-)$ represents the target Q-value, and $\theta_i$ and $\theta_i^{-}$ are the weights of the Q-network and target network, respectively, at iteration $i$.

Two critical innovations improve the stability and convergence of DQN. First, the use of \emph{experience replay} allows the agent to store past transitions $(s_t, a_t, r_t, s_{t+1})$ in a replay buffer $D = \{e_1, \dots, e_N\}$, which are then randomly sampled during training. This helps break the temporal correlation between consecutive samples and reduces variance. Second, a \emph{target network} is used to stabilize the learning process. The target network is a copy of the Q-network, and its weights are updated every $C$ step. By holding the target network constant for several updates, DQN reduces the risk of divergence and policy oscillations that can occur when the target values change too rapidly.

The full pseudocode for the DQN algorithm is provided in Algorithm \ref{alg:1}, adapted from \citep{mnih2015human}.
\begin{algorithm}[!ht]
\caption{DQN}\label{alg:1}
\begin{algorithmic}[1]
\State Initialize replay memory $D$ to size $N$
\State Initialize Q-network $Q$ with random weights $\theta$
\State Initialize target network $\hat{Q}$ with weights $\theta^-=\theta$
\For{$\text{episode}=1,\dots,M$}
    \State Initialize state $s_0$
    \For{$t=0,1,\dots,T-1$}
        \State With probability $\varepsilon$, select random action $a_t$
        \State Otherwise, select $a_t=\text{argmax}_a Q(s_t,a_t;\theta)$
        \State Execute $a_t$ and observe $r_t$ and $s_{t+1}$
        \State Store transition $(s_t, a_t, r_t, s_{t+1})$ in $D$
        \State Sample transitions $(s_j,a_j,r_j,s_{j+1})$ from $D$
        \State Set $y_j=r_j$ if episode ends at step $j+1$
        \State Otherwise, set $y_j=r_j + \gamma \max\limits_{a'} \hat{Q}(s_{j+1}, a'; \theta^-)$
    \State Do gradient descent on $\left( y_j-Q\left( s_j,a_j;\theta \right)\right)^2$
    \State Every $C$ steps, set $\hat{Q}=Q$
    \EndFor
\EndFor
\State \textbf{Output:} Q-network weights $\theta$
\end{algorithmic}
\end{algorithm}

For our numerical experiments, we use the DQN implementation from \citep{wu2024new}. More details are given in Section \ref{sec:5.2}.

\subsection{PPO algorithm}
Proximal Policy Optimization (PPO) is based on the actor-critic method, Trust Region Policy Optimization (TRPO) \citep{schulman2015trust}, which seeks to solve
\begin{equation} \label{eq:9}
\begin{split}
\underset{\theta}{\text{max}}& \hspace{0.25cm}\mathbb{E}_{t}\left[ \frac{\pi_{\theta}(a_t|s_t)}{\pi_{\theta_\text{old}}(a_t|s_t)}A_t \right]\\
\text{s.t.}&\hspace{0.25cm} \mathbb{E}_t\left[ \text{KL} \left( \pi_{\theta_\text{old}}(\cdot|s_t)||\pi_{\theta}(\cdot|s_t) \right) \right]\leq \delta,
\end{split}
\end{equation}
where $A_t=A^{\pi_{\theta_{\text{old}}}}(s_t,a_t)$ is the advantage function, defined as $A^{\pi}(s_t,a_t)=Q^{\pi}(s_t,a_t)-V^{\pi}(s_t)$, $\text{KL}\left(\cdot||\cdot \right)$ is the KL divergence function, and $\delta$ is a hyperparameter \citep{schulman2015trust}. This problem is solved by approximating the advantage function $A_t$ and the KL constraint, followed by the conjugate gradient method. TRPO finds large policy updates while ensuring the updates are constrained by the KL divergence to prevent the policy from deviating too far from the previous one.

PPO addresses the need for the quadratic approximation in TRPO’s KL constraint by introducing a clip function in the loss:
\begin{equation}
\label{eq:10}
\begin{aligned}
    L^{\textrm{CLIP}}_t(\theta)= \mathbb{E}_t\left[ \min \left( \frac{\pi_{\theta}(a_t|s_t)}{\pi_{\theta_\text{old}}(a_t|s_t)}A_t, \right. \right. & \\
    \left. \left. \text{clip}\left( \frac{\pi_{\theta}(a_t|s_t)}{\pi_{\theta_\text{old}}(a_t|s_t)}, 1-\varepsilon, 1+\varepsilon \right)A_t \right) \right].
\end{aligned}
\end{equation}
Here, the clip function keeps the ratio of the new policy to the old policy within the range $[1-\varepsilon, 1+\varepsilon]$, preventing the policy from updating too aggressively and eliminating the need for an explicit optimization constraint \citep{schulman2017proximal}. If we share parameters between the policy network and the value function network, we can modify $L_t^{\textrm{CLIP}}$ by adding a value function error term and an entropy term:
\begin{equation}\label{eq:11}
    L_t(\theta)= \mathbb{E}_t\left[ L^{\textrm{CLIP}}_t(\theta) - c_1\left(V^{\pi_\theta}(s_t)-V_t^{\text{targ}} \right)^2+c_2S[\pi_\theta](s_t) \right].
\end{equation}
In (\ref{eq:11}), $c_1$ and $c_2$ are hyperparameters, and $S$ is an entropy function. The full pseudocode for PPO with shared networks is shown in Algorithm \ref{alg:2} \citep{schulman2017proximal}.

\begin{algorithm}[!ht]
\caption{PPO}\label{alg:2}
\begin{algorithmic}[1]
\State Initialize parameters $\theta_\text{old}$
\For{$\text{iteration}=1,2,\dots$}
    \For{$\text{actor}=1,2,\dots,N$}
        \State Run $\pi_{\theta_\text{old}}$ for $T$ timesteps
        \State Compute advantage functions$A_1,\dots,A_T$
    \EndFor
    \State Optimize $L$ with $K \text{epochs}, \text{minibatch size}\leq NT$
    \State Set $\theta_{\text{old}}=\theta$
\EndFor
\State \textbf{Output:} Network weights $\theta$
\end{algorithmic}
\end{algorithm}

To implement PPO, we use the PPO package from \texttt{Stable-Baselines3 1.6.0} \citep{stable-baselines3}. More details are given in Section \ref{sec:5.2}.

\section{gym-DSSAT}
\label{sec:gym-dssat}
In this section, we outline the crop management process, which includes the simulator as well as the fertilization, irrigation, and combined fertilization and irrigation problems.

\subsection{Environment details} \label{sec:4.1}
This paper utilizes the gym-DSSAT environment, an open-source RL framework built with the popular Python RL toolkit OpenAI Gym\footnote{\url{https://www.gymlibrary.dev/index.html}} and based on the Decision Support System for Agrotechnology Transfer (DSSAT) crop simulator \citep{gautron2022gym}. The default configuration of gym-DSSAT simulates a maize experiment conducted in 1982 at the University of Florida farm in Gainesville, Florida, USA. Each time step within the simulation corresponds to one day, with a full episode representing an entire growing season of approximately 160 days. The episode concludes either when the crops are ready to be harvested or when a crop failure occurs. Additionally, weather conditions are stochastically generated for each episode, adding variability to the simulation. The gym-DSSAT environment provides three distinct management modes: fertilization, irrigation, and a combined mode that allows both fertilization and irrigation. In each mode, the RL agent interacts with the environment and makes decisions on a daily basis. These modes reflect real-world agricultural challenges where farmers must decide how much nitrogen fertilizer to apply, how much water to use for irrigation, or how to balance both inputs. Detailed descriptions of these modes are provided in Sections \ref{sec:4.2} - \ref{sec:4.4}.

To mirror the real-world uncertainty that farmers face, the gym-DSSAT environment limits the agent's access to the full set of state variables available in DSSAT, making only a subset of variables observable at each time step. This design introduces the problem of partial observability, where the RL agent must make decisions based on incomplete information. Despite this partial observability, gym-DSSAT can still be effectively treated as a Markov Decision Process (MDP), with no significant complications in implementation. A comprehensive description of the observation space for each mode can be found in \citep{gautron2022gym}.

\subsection{Fertilization problem} \label{sec:4.2}
In the fertilization problem, the RL agent is tasked with determining the daily application of nitrogen fertilizer (kg/ha) within the range of $[0, 200]$. During this task, the crops are strictly rainfed, meaning the agent has no control over irrigation, and planting operations are automatically managed by DSSAT based on soil temperature and humidity conditions. The primary goal is to maximize the nitrogen uptake by the crops while minimizing excessive fertilizer use.

The reward function for the fertilization problem encourages efficient nitrogen use by balancing nitrogen uptake with the cost of fertilization. Specifically, let $\text{trnu}(t)$ denote the plant's nitrogen uptake between day $t$ and $t+1$, and let $N_t$ represent the amount of nitrogen applied on day $t$. The reward function is defined as
\begin{equation}\label{eq:12} r_F(t) = \text{trnu}(t, t+1) - 0.5 N_t, \end{equation}
where the first term rewards nitrogen uptake, and the second term penalizes excessive fertilization from the previous day. This reward structure reflects the trade-off between promoting plant growth and avoiding unnecessary fertilizer application \citep{gautron2022gym}.

\subsection{Irrigation problem} \label{sec:4.3}
The irrigation problem focuses on the daily application of water (L/$\text{m}^2$) within the range of $[0, 50]$. The agent is responsible for managing irrigation, while fertilization is handled deterministically, with small amounts of nitrogen fertilizer applied automatically at 40, 45, and 80 days after planting. The agent must optimize water usage to maximize crop growth, balancing the water provided against the plant's biomass gain.

Let $\text{topwt}(t, t+1)$ represent the change in above-ground biomass (kg/ha) between day $t$ and $t+1$, and let $W_t$ denote the amount of water applied on day $t$. The reward function for the irrigation problem is defined as
\begin{equation}\label{eq:13} r_I(t) = \text{topwt}(t, t+1) - 15 W_t, \end{equation}
where the first term rewards increases in biomass, and the second term penalizes excessive water use. This reward structure encourages the agent to apply water judiciously, promoting crop growth while minimizing over-irrigation.

\subsection{Mixed problem} \label{sec:4.4}
The mixed problem combines the tasks of fertilization and irrigation, requiring the RL agent to simultaneously manage both nitrogen application and water usage. The agent can apply nitrogen fertilizer in the range of $[0, 200]$ kg/ha and provide water for irrigation within the range of $[0, 50]$ L/m² each day. The observation space in this mode is the union of the observation spaces from the fertilization and irrigation problems.

The reward at each interaction is a vector $(r_F(t), r_I(t))$, representing the rewards for fertilization and irrigation. However, since many RL algorithms, such as PPO and DQN, require scalar rewards, a scalar reward function is used for numerical experiments, as suggested by \citep{tao2022optimizing}. The reward function for the mixed problem is defined as
\begin{equation}\label{eq:14}
r_M(t) = 
\begin{cases}
    w_1Y-w_2 N_t-w_3 W_t - w_4N_{l,t} & \text{if harvest at } t,\\
    -w_2 N_t-w_3 W_t - w_4N_{l,t} & \text{otherwise},
\end{cases}
\end{equation}
where $w_1, w_2, w_3,$ and $w_4$ are weights that represent the importance of yield ($Y$), nitrogen usage, water usage, and nitrate leakage ($N_{l,t}$), respectively. These weights can be adjusted depending on the priorities of the management task. For instance, when $w_1 = 0.158, w_2 = 0.79, w_3 = 1.1$, and $w_4 = 0$, the reward function balances the competing objectives of maximizing yield and minimizing resource use.

The mixed problem presents the agent with a more complex decision-making environment, where it must balance the trade-offs between nitrogen and water application to maximize yield while minimizing negative environmental impacts, such as nitrate leakage. This combined task closely mirrors real-world agricultural management, where farmers must simultaneously manage multiple inputs to optimize crop production outcomes.

\section{Numerical Experiments}
\label{sec:experiments}
In this section, we compare the performance of two RL algorithms, DQN and PPO, in the gym-DSSAT environment. Our focus is on assessing how these methods perform on fertilization, irrigation, and a mixed strategy that incorporates both fertilization and irrigation policies.

\subsection{Experimental setup}
The experiments were conducted in gym-DSSAT's default environment, simulating the 1982 maize experiment at the University of Florida, as mentioned in Section \ref{sec:4.1}. Both RL algorithms—PPO and DQN—were trained to learn three distinct policies: a fertilization policy, an irrigation policy, and a mixed policy combining both fertilization and irrigation. The performance of these learned policies was compared against two baseline methods previously used in \citep{gautron2022gym}:
\begin{itemize}
    \item \textit{Null policy}: This baseline involves applying no nitrogen fertilizer (0 kg/ha) and no irrigation water $0~\text{L}/\text{m}^2$. The Null policy serves as a control to evaluate crop growth with only natural inputs, such as nitrogen from the soil and rainwater. This comparison is essential to determine the effect of RL-based intervention \citep{vanlauwe2011agronomic, howell2003irrigation}.
    \item \textit{Expert policy}: The Expert policy replicates the original fertilization strategy from the 1982 experiment and approximates the irrigation schedule. Fertilization and irrigation are applied deterministically based on the number of days after planting, representing the traditional agronomic approach \citep{hunt1998data}.
\end{itemize}

It is important to note that these baseline methods differ from those used in other RL studies such as \citep{tao2022optimizing}, where fertilizers and irrigation were applied based on crop growth indicators \citep{wright2022field}. For our fertilization and irrigation tasks, we used reward functions (\ref{eq:12}) and (\ref{eq:13}), while for the mixed experiment, we applied the reward function (\ref{eq:14}), configured to prioritize economic profit. Each trained RL policy was tested against these baselines over 1000 episodes, with different stochastic weather conditions to ensure robustness. The main evaluation metric was the average cumulative reward, which allowed us to assess overall policy performance under varying environmental conditions.

\subsection{Implementation details} \label{sec:5.2}
To implement PPO, we used the \texttt{Stable-Baselines3 1.6.0} library \citep{stable-baselines3}, with default hyperparameters. The discount factor was set to $\gamma = 1.0$, consistent with the gym-DSSAT environment settings in \citep{gautron2022gym}. PPO was trained for $10^6$ time steps, with validation checks every 1000 steps. During evaluation, the actions were selected stochastically for the fertilization and irrigation tasks. However, for the mixed problem, better results were achieved by using deterministic action selection, a behavior controlled within the \texttt{predict} function of the PPO implementation.

For DQN, we utilized the Python implementation from \citep{wu2024new} \footnote{\url{https://github.com/jingwu6/LM_AG/tree/main}}, which supports the discretization of gym-DSSAT's continuous action space. Specifically, the action space was discretized as follows:
\begin{equation} \label{eq:15}
    A=\{40i \text{ kg/ha } \text{N fertilizer } \& \hspace{0.1cm} 6j \text{ L/$\text{m}^2$} \text{ irrigation water}\},
\end{equation}
where $i, j = 0, 1, 2, 3, 4$. The DQN model was trained for 4000 episodes with a discount factor of $\gamma = 0.99$, using the $\varepsilon$-greedy method for action selection. The neural network architecture consisted of three hidden layers, each with 256 units. The learning rate was set to $10^{-5}$, and the batch size was 1024.

\subsection{Policy training}
\begin{figure}[ht]
\centering
\includegraphics[width=0.45\textwidth]{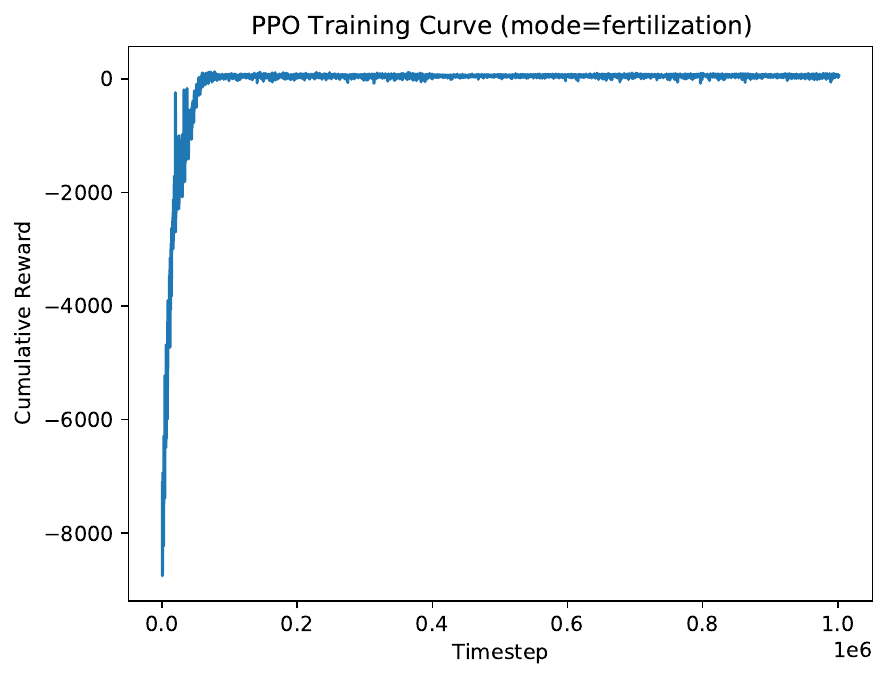}
\includegraphics[width=0.45\textwidth]{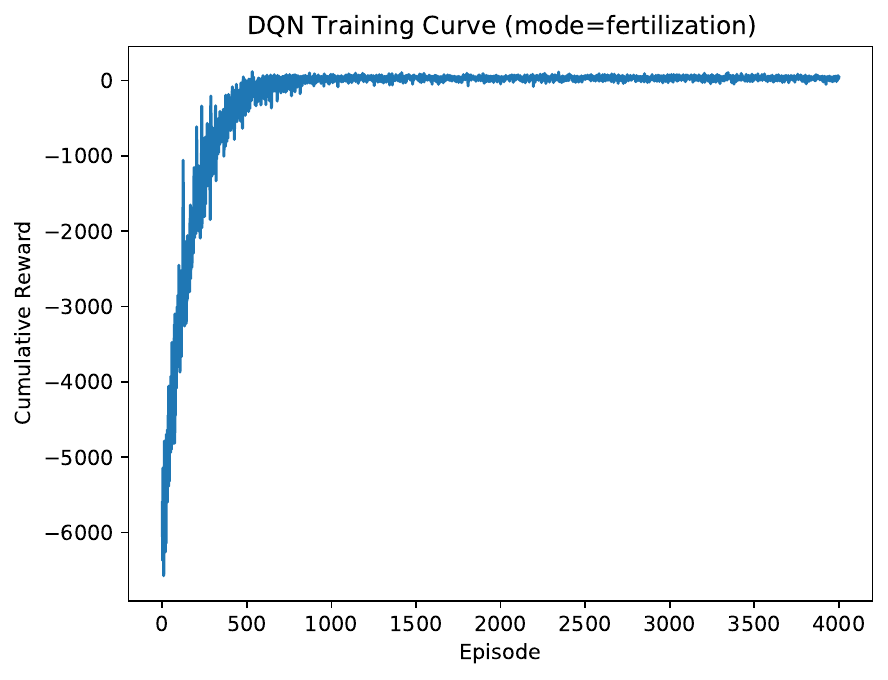}
\caption{PPO and DQN training curves for the fertilization problem. The horizontal axis measures training iterations and the vertical axis measures cumulative rewards.}
\label{fig:figure1}
\end{figure}

\begin{figure}[ht]
\centering
\includegraphics[width=0.45\textwidth]{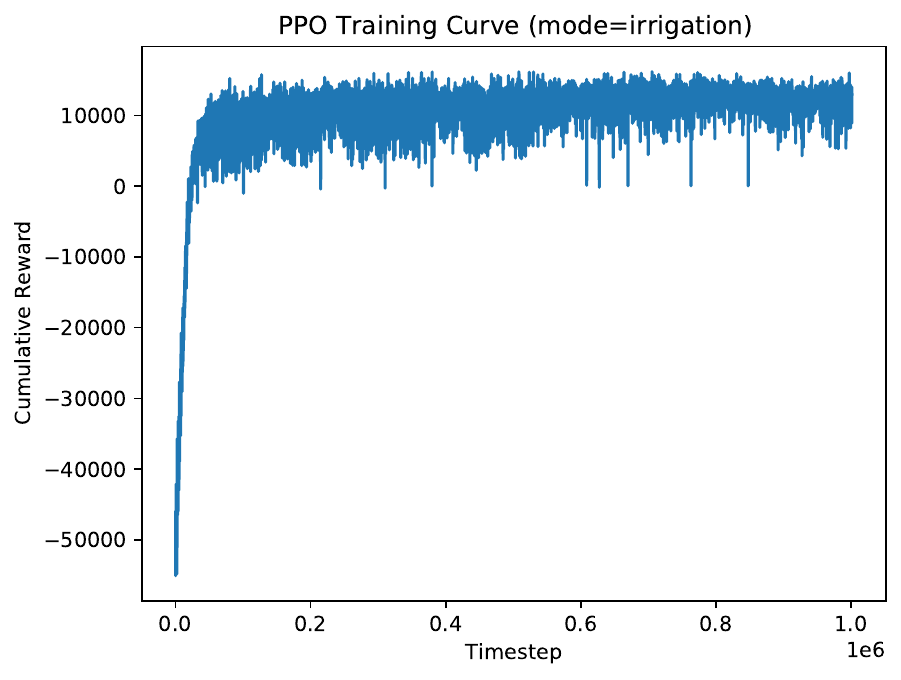}
\includegraphics[width=0.45\textwidth]{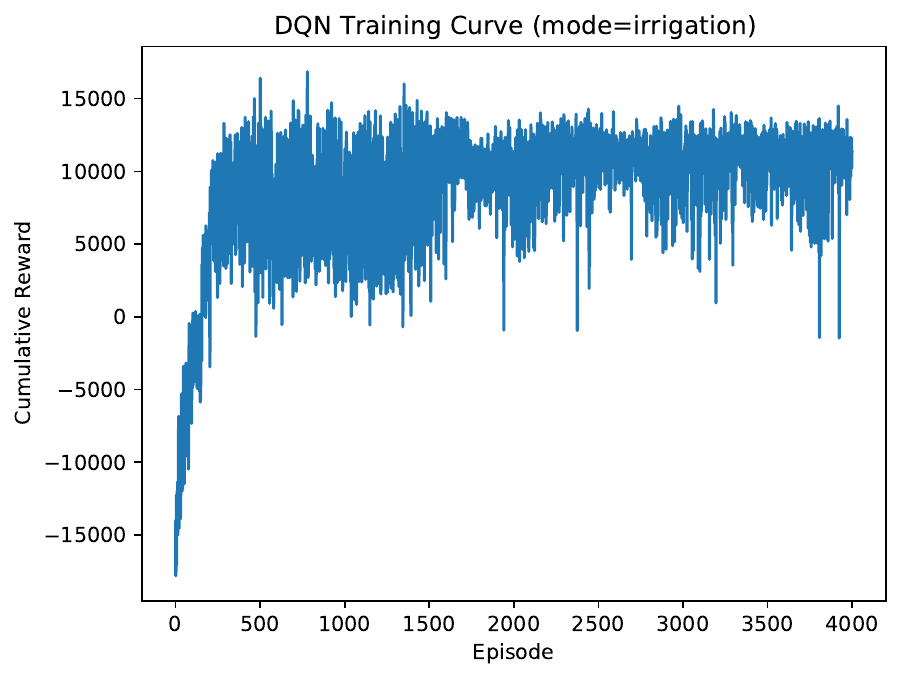}
\caption{PPO and DQN training curves for the irrigation problem. The horizontal axis measures training iterations and the vertical axis measures cumulative rewards.}
\label{fig:figure2}
\end{figure}

\begin{figure*}[ht]
\centering
\includegraphics[width=0.45\textwidth]{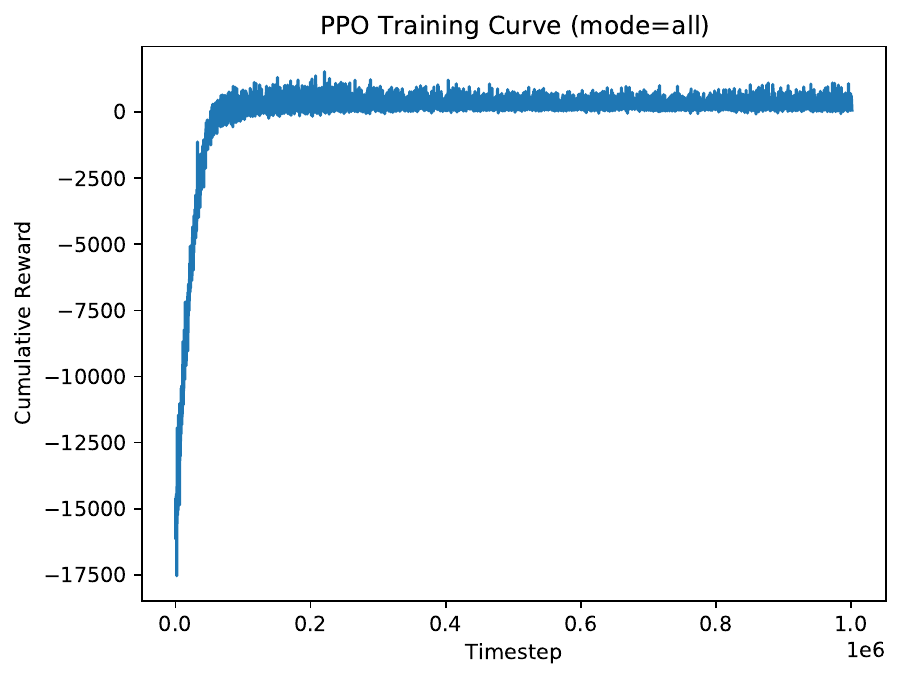}
\includegraphics[width=0.45\textwidth]{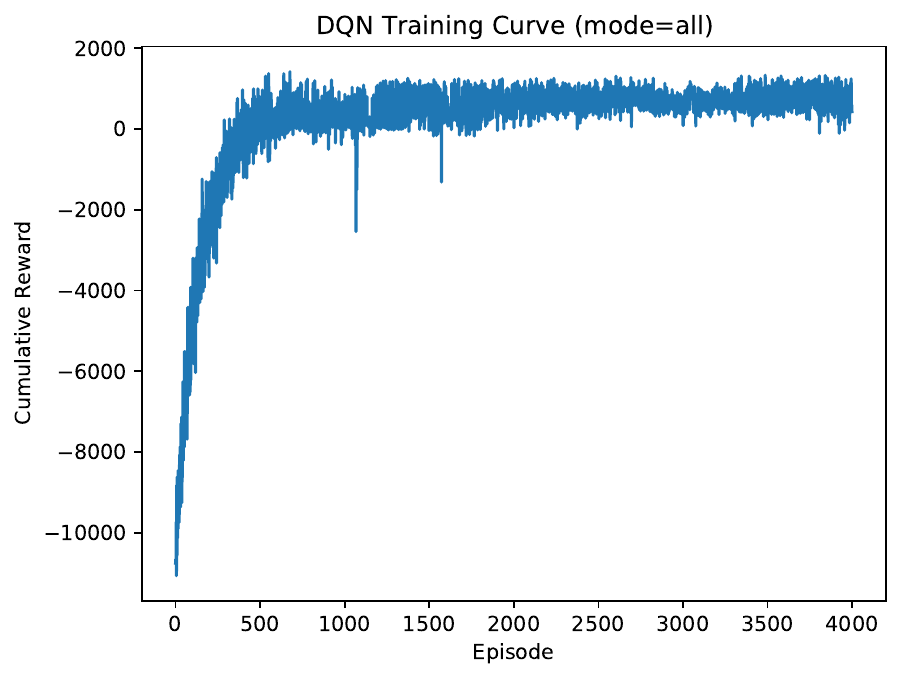}
\caption{PPO and DQN training curves for the mixed problem. The horizontal axis measures training iterations and the vertical axis measures cumulative rewards.}
\label{fig:figure3}
\end{figure*}

The training curves for PPO and DQN across the three tasks—fertilization, irrigation, and mixed—are shown in Figures \ref{fig:figure1}, \ref{fig:figure2}, and \ref{fig:figure3}, respectively. These curves plot the cumulative rewards as a function of the number of training iterations.

For the fertilization problem, as shown in Figure \ref{fig:figure1}, both PPO and DQN converged early in the training process, indicating that learning an optimal fertilization policy is relatively quick for both algorithms. The steady rise in cumulative rewards suggests that both methods quickly learned how to optimize nitrogen application to enhance crop growth.

In contrast, Figure \ref{fig:figure2} shows more oscillation during training for the irrigation problem, particularly for DQN. These oscillations indicate instability in the learning process, potentially due to the increased complexity of water management compared to fertilization. The oscillations suggest that irrigation policies are more difficult for both PPO and DQN to learn effectively, which may be due to the stochastic nature of rainfall in the simulation and the more delicate balance required to optimize water usage.

The mixed problem, depicted in Figure \ref{fig:figure3}, demonstrates slower convergence for both PPO and DQN compared to the fertilization task. While the oscillations were less severe than those seen in the irrigation problem, DQN exhibited more fluctuations in cumulative rewards than PPO, though it ultimately achieved higher rewards. This suggests that while DQN is more sensitive to parameter tuning and experiences more variability during training, it can outperform PPO when the model is properly optimized.

\subsection{Evaluation results}
Table~\ref{tab:Table 1} summarizes the average cumulative rewards and their standard deviations over 1000 test episodes for each policy (Null, Expert, PPO, and DQN) across the three RL problems: fertilization, irrigation, and the mixed (all) problem. The performance of each policy is measured by the average cumulative reward, with the highest score bolded.

For the fertilization problem, PPO outperformed both the Expert and DQN policies, achieving the highest average cumulative rewards with lower variance than the Expert policy. DQN performed poorly on this task, with the lowest average cumulative rewards and the largest variance, indicating a less consistent learning process. This suggests that PPO was more efficient in learning a fertilization strategy that balanced nitrogen uptake with penalizing excessive fertilizer use.

In the irrigation problem, PPO again achieved the highest cumulative rewards, though it exhibited much higher variance compared to the Expert policy. This high variance may indicate that while PPO found effective irrigation strategies, its performance fluctuated due to the stochastic nature of rainfall in the environment. DQN, while closer in performance to PPO, still trailed the Expert policy and showed the highest variance, suggesting that it struggled to consistently balance water usage across episodes.

For the mixed problem (fertilization and irrigation), the Expert policy achieved the highest average cumulative rewards. DQN followed closely, outperforming PPO by a significant margin. Interestingly, PPO performed much worse in this combined task compared to its performance in the single-task problems (fertilization and irrigation). Although PPO managed to outperform the Null policy, its poor performance in the mixed problem may be attributed to its difficulty in managing the complexity of simultaneously optimizing both nitrogen and water applications.

\begin{figure*}[ht]
\centering
\includegraphics[width=0.32\textwidth]{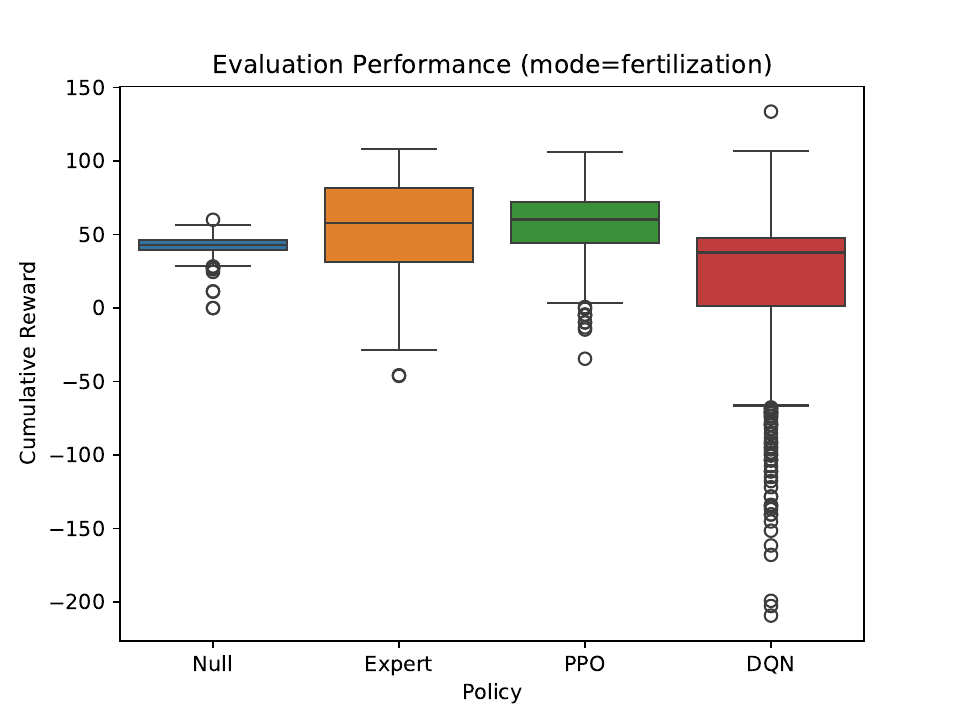}
\includegraphics[width=0.32\textwidth]{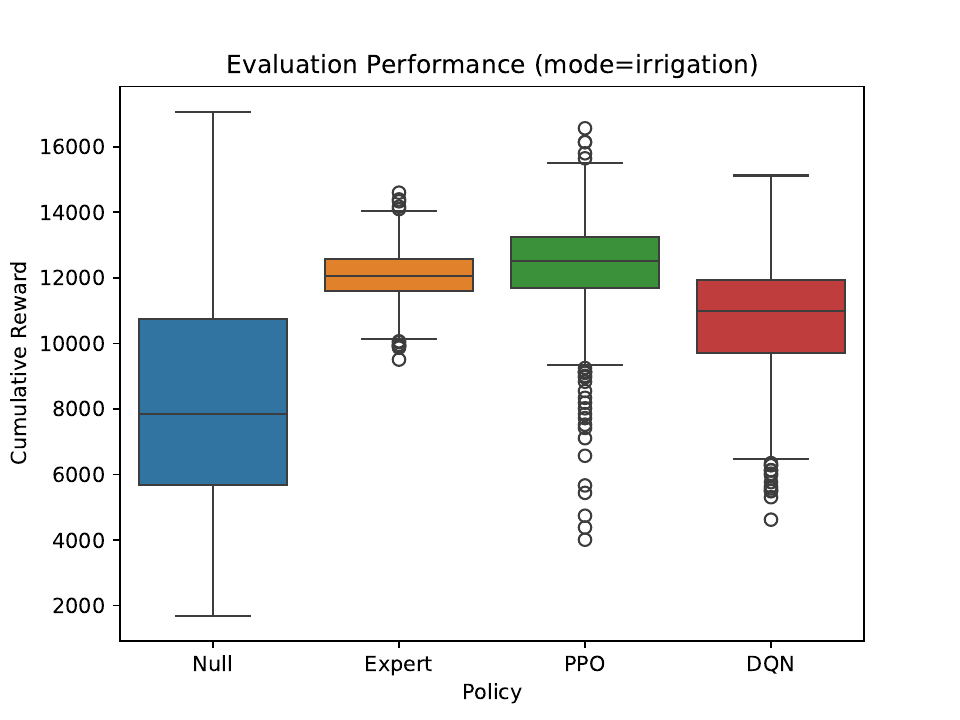}
\includegraphics[width=0.32\textwidth]{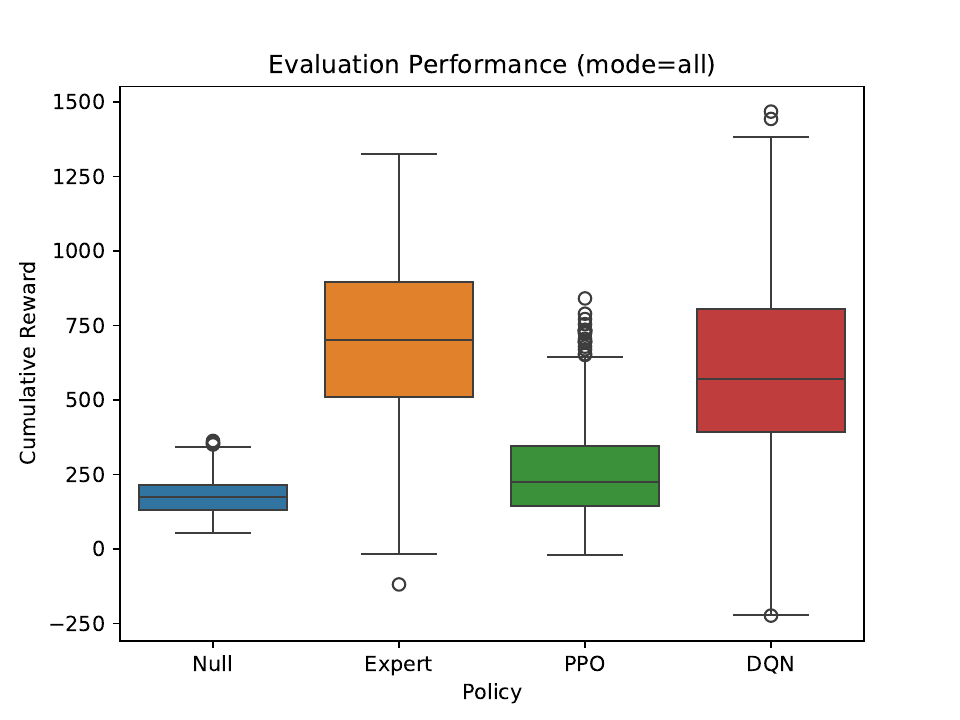}
\caption{Evaluation results shown as box plots. The vertical axis measures cumulative rewards for the 1000 test episodes.}
\label{fig:figure4}
\end{figure*}

Figure \ref{fig:figure4} presents these evaluation results in the form of box plots, providing a visual comparison of the cumulative rewards for each policy across the three tasks. The box plots reveal the distribution of rewards across the 1000 test episodes, highlighting the performance consistency and variability of each approach.

\begin{table*}[ht]
\caption{Average cumulative rewards and standard deviations over 1000 test episodes. The highest scores are bolded.}
    \small
    \centering
    \begin{tabular}{|c||c|c|c|c|}
    \hline
       & Null & Expert & PPO & DQN\\
       \hline
      Fertilization & $42.52\pm5.87$    &$55.24\pm30.12$ &   $\textbf{57.0}\pm20.6$&$21.23\pm45.81$\\
      Irrigation & $8213.84\pm3191.97$  & $12068.41\pm750.9$   &$\textbf{12389.76}\pm1379.06$&$10765.19\pm1700.18$\\
      All &$175.52\pm59.55$ & $\textbf{691.87}\pm 272.6$&  $257.09\pm 149.12$&$594.88\pm 299.17$\\
      \hline
    \end{tabular}
    \label{tab:Table 1}
\end{table*}


\begin{figure*}[ht]
\centering
\begin{tabular}{cc}
    \includegraphics[width=0.45\textwidth]{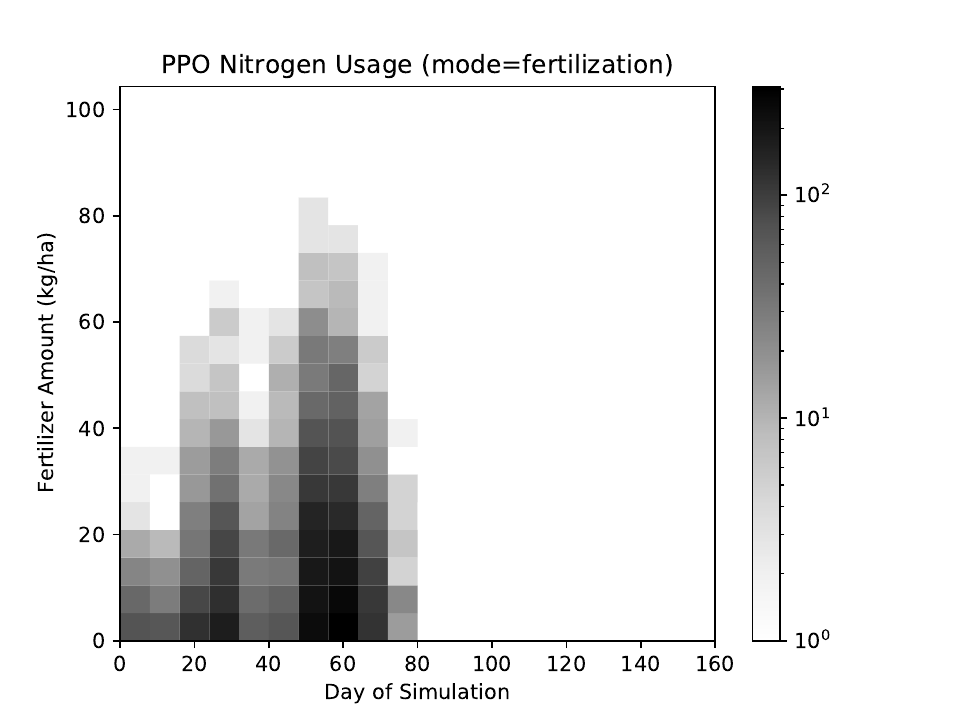} &
    \includegraphics[width=0.45\textwidth]{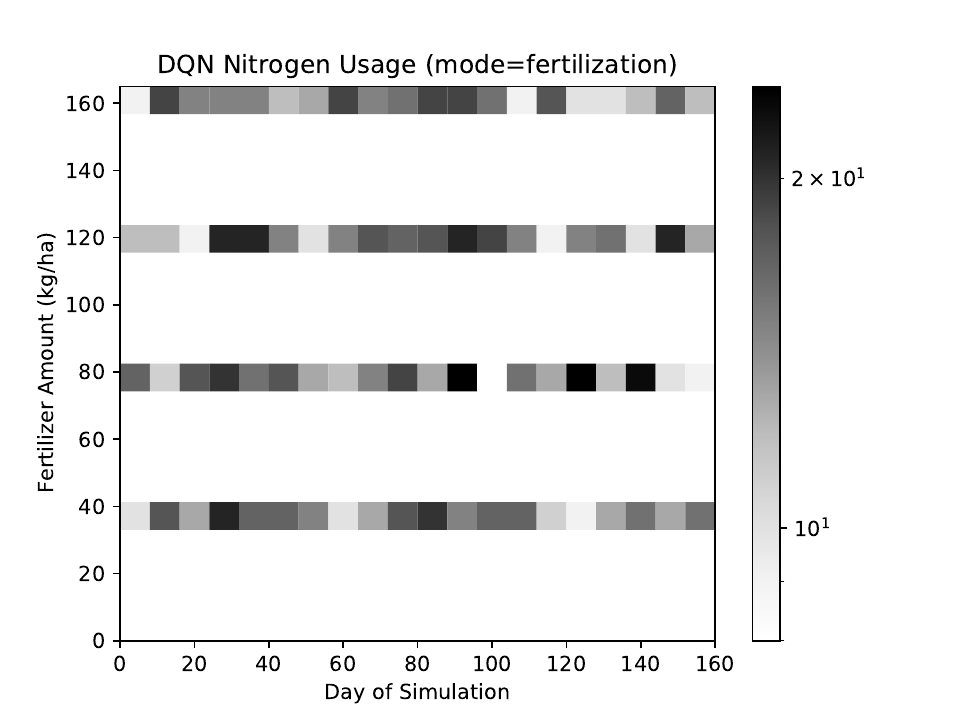} \\
    \includegraphics[width=0.45\textwidth]{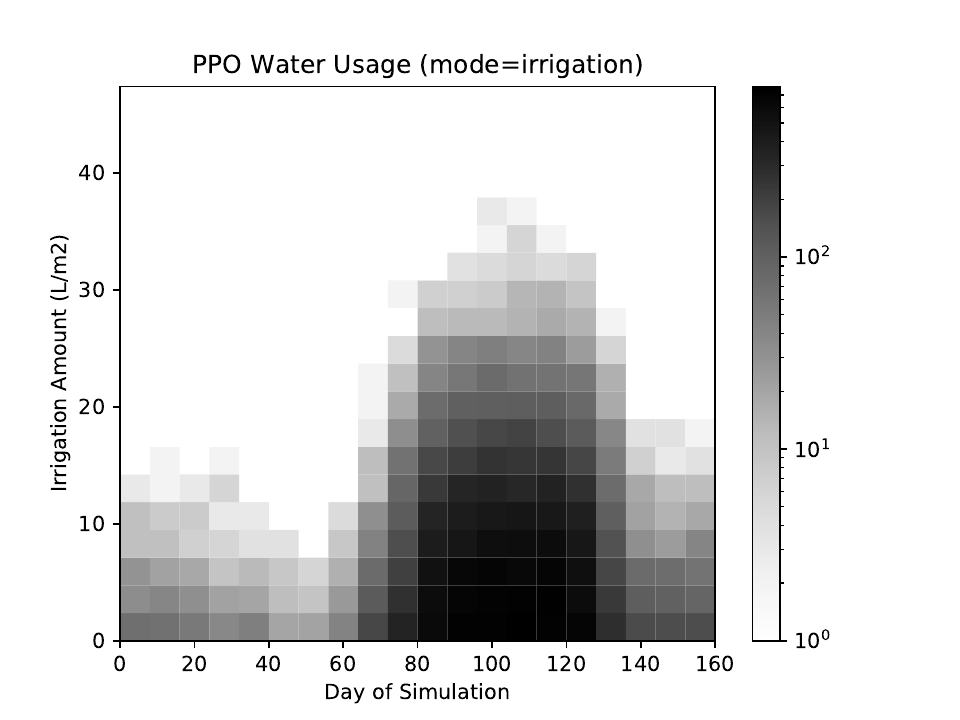} &
    \includegraphics[width=0.45\textwidth]{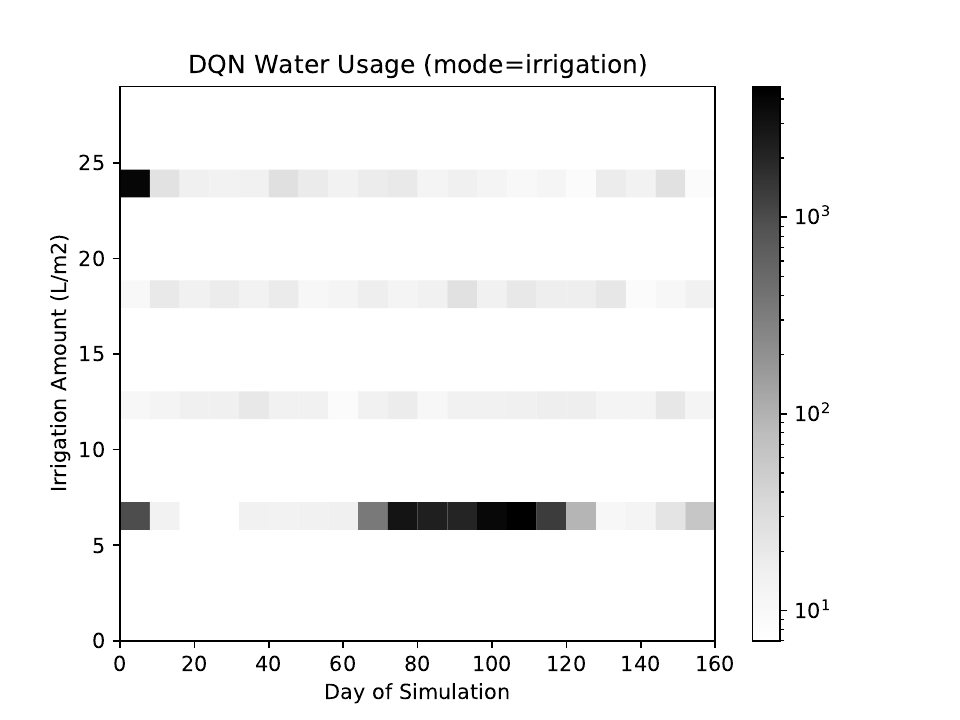} \\
\end{tabular}
\caption{2D histograms showing the frequency of nonzero fertilization and irrigation applications during testing for the fertilization problem and the irrigation problem. Darker areas correspond to higher frequencies.}
\label{fig:figure5}
\end{figure*}

\begin{figure*}[ht]
\centering
\begin{tabular}{cc}
    \includegraphics[width=0.45\textwidth]{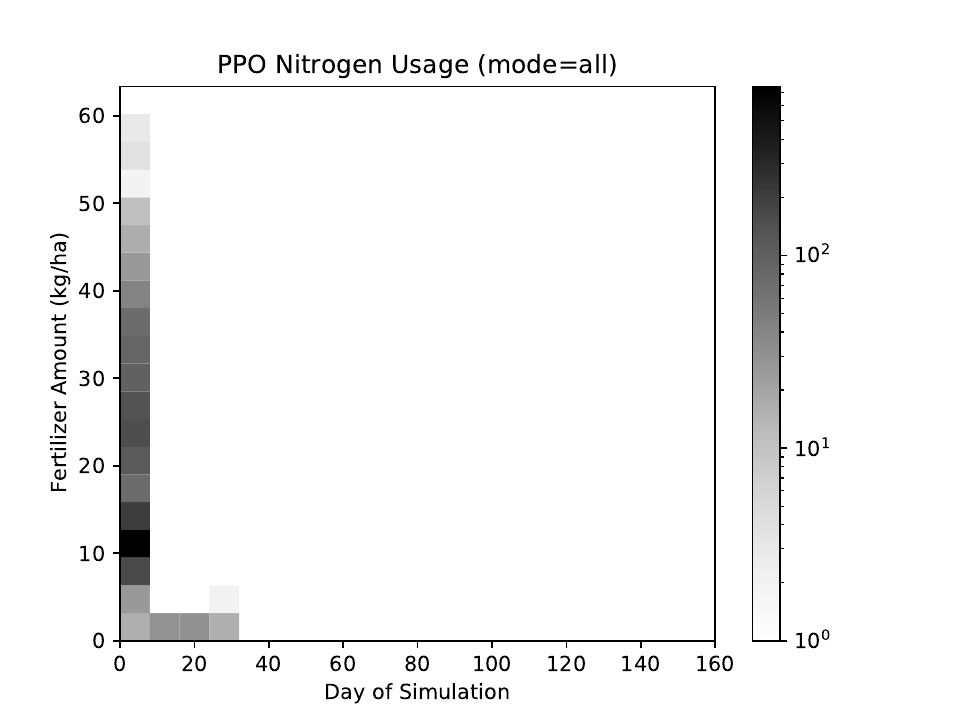} &
    \includegraphics[width=0.45\textwidth]{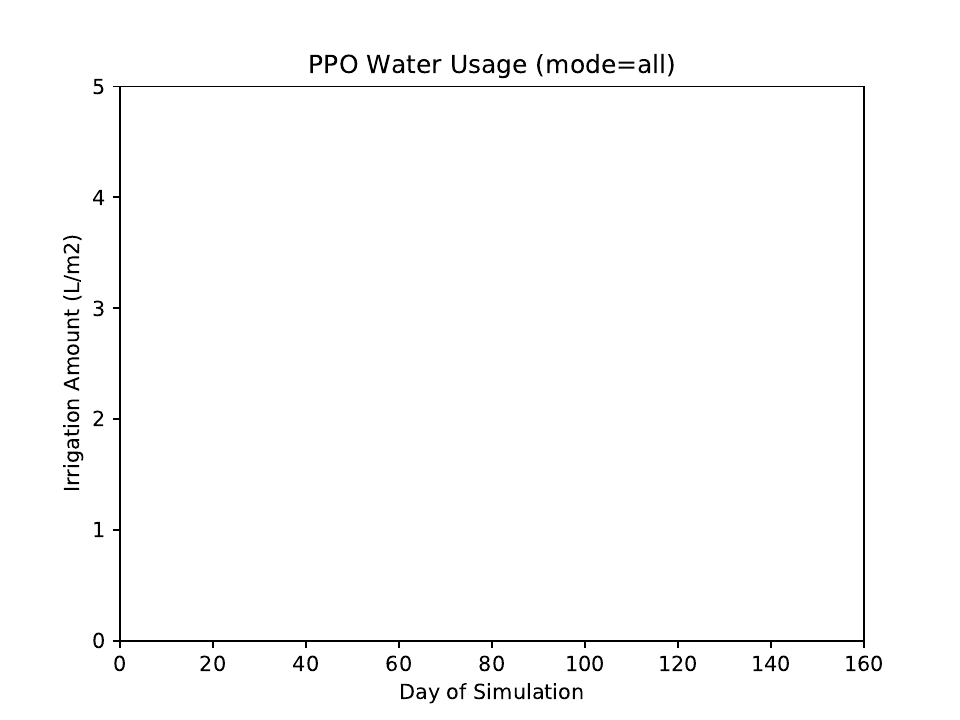} \\
    \includegraphics[width=0.45\textwidth]{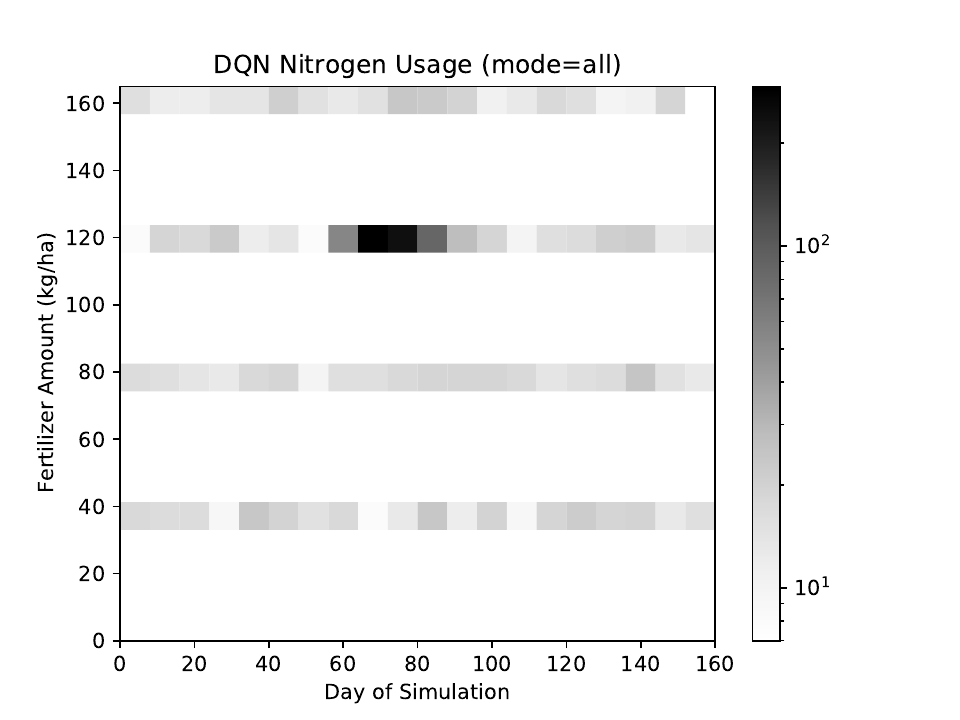} &
    \includegraphics[width=0.45\textwidth]{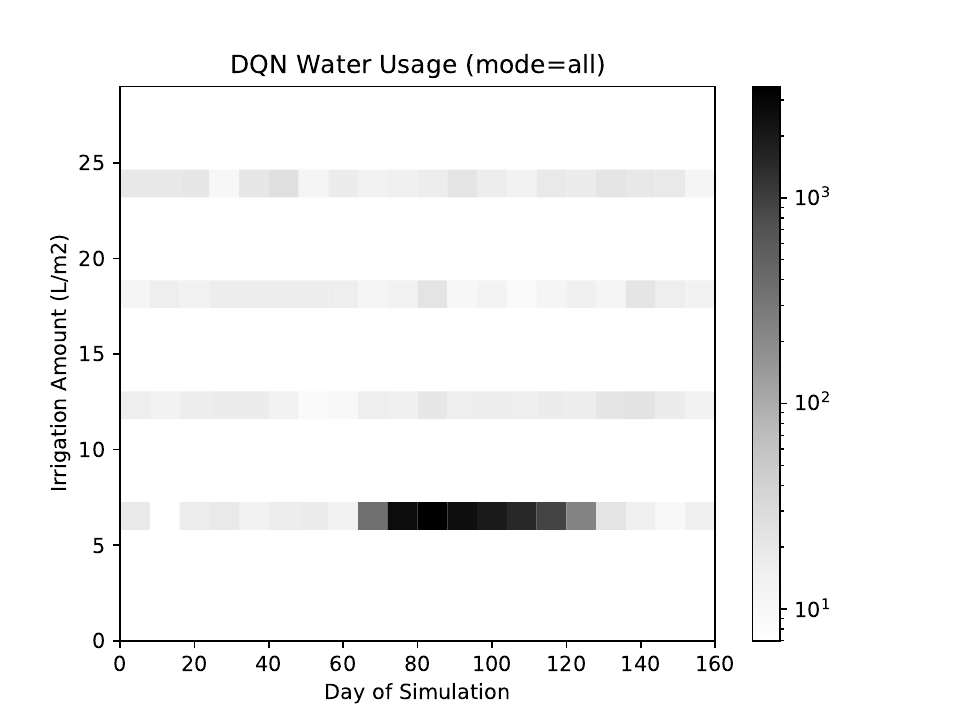} \\
\end{tabular}
\caption{2D histograms showing the frequency of nonzero fertilization and irrigation applications during testing for the mixed problem. Note, the PPO water usage histogram is empty because PPO provided no irrigation throughout all test episodes.}
\label{fig:figure6}
\end{figure*}

Figures \ref{fig:figure5} and \ref{fig:figure6} present 2D histograms that show the frequency of nonzero fertilization and irrigation applications across all test episodes. These histograms offer insights into how the RL agents applied nitrogen and water throughout the growing season. For the fertilization problem (Figure \ref{fig:figure5}), PPO concentrated most of its nitrogen applications around day 60, typically applying between 0 and 50 kg/ha of nitrogen. In contrast, DQN displayed no clear pattern, with nitrogen applications scattered across the growing season, likely contributing to its lower performance. This lack of a focused strategy likely explains why DQN struggled to outperform PPO and the Expert policy in this task. In the irrigation problem, both PPO and DQN applied water primarily between days 80 and 120. PPO tended to apply moderate amounts of water (0-20 L$/\text{m}^2$), whereas DQN consistently applied smaller amounts of water (around 6 L$/\text{m}^2$). The uniformity of DQN’s irrigation application may have limited its ability to adapt to stochastic weather conditions, resulting in its lower performance relative to PPO.

For the mixed problem (Figure \ref{fig:figure6}), PPO applied nitrogen only in the first few days after planting and surprisingly applied no water throughout the entire growing season. This extreme sparsity in its actions explains why PPO performed similarly to the Null policy in this task. On the other hand, DQN applied nitrogen primarily between days 60 and 80 at a high rate (around 120 kg/ha), and it applied irrigation water consistently between days 80 and 120. DQN’s more balanced approach to managing both fertilization and irrigation contributed to its better performance relative to PPO in the mixed problem.

In summary, PPO demonstrated superior performance in single-task problems like fertilization and irrigation, but struggled in the more complex mixed problem. On the other hand, DQN exhibited higher variance but performed better in the mixed task by employing a more balanced application strategy. These results suggest that while PPO may be better suited for simpler management tasks, DQN may have an advantage in scenarios that require simultaneous optimization of multiple variables.

\section{Discussion}
The goal of this study was to evaluate the performance of PPO and DQN in the gym-DSSAT environment using default settings and parameters. Our numerical experiments showed that PPO outperformed DQN on individual tasks such as fertilization and irrigation, but it performed worse than DQN on the mixed task, which required managing both fertilization and irrigation simultaneously. Notably, PPO failed to make nonzero irrigation decisions in the mixed task, leading to a performance close to that of the Null policy. This behavior could be attributed to suboptimal parameter selection, particularly in scenarios requiring simultaneous optimization across multiple inputs. One potential solution would be to introduce parameter selection or hyperparameter tuning as part of the PPO training phase to optimize its performance across more complex tasks.

Suboptimal parameter choices also seem to have affected DQN’s performance, particularly in the fertilization problem, where it failed to surpass the Expert policy in cumulative rewards. This indicates that further tuning or adjustments to the reward structure could help improve DQN's ability to generalize across different conditions. Moreover, the lack of robust validation procedures during model selection may have contributed to both algorithms’ inconsistent performances. Future experiments should incorporate systematic validation techniques, such as cross-validation or grid search, to ensure that the models are evaluated using optimal configurations. Additionally, future studies should investigate the impact of different reward functions and performance criteria to better understand how these factors influence learning outcomes. Testing alternative reward formulations, such as more detailed economic or environmental metrics, may yield richer insights into the performance of RL algorithms in agricultural management. Overall, the mixed results between PPO and DQN highlight the need for more adaptive algorithms that can efficiently handle both single-task and multi-task learning scenarios in crop management.

Despite their potential, RL algorithms like PPO and DQN present several shortcomings, particularly in agricultural applications such as crop management \citep{levine2020offline}. First, RL algorithms require extensive interaction with the environment to learn optimal policies, which in real-world agricultural settings would demand a significant amount of time, resources, and data. This is impractical for real farms, where each interaction corresponds to an entire growing season, making it difficult to simulate enough real-time data to train the models effectively. Moreover, RL algorithms are sensitive to hyperparameters, and optimizing these parameters typically requires a trial-and-error approach, leading to inefficient training. Furthermore, RL models may struggle to generalize when confronted with variability in environmental conditions (e.g., changes in weather, soil conditions) that were not encountered during training, making them less robust in practical scenarios \citep{chen2024learning}.

To address some of these shortcomings, offline reinforcement learning (offline RL) offers a promising alternative. Unlike traditional RL, which requires continuous interactions with the environment, offline RL learns from static datasets that are collected in advance. These datasets can consist of historical data from previous growing seasons, sensor data, and expert decisions, making offline RL much more feasible for agriculture. Since farmers often have access to large amounts of historical data, offline RL eliminates the need for costly and time-consuming real-time interactions. Additionally, offline RL allows algorithms to learn from past mistakes and successes, potentially leading to more robust and well-informed decision-making processes \citep{levine2020offline}.

\section{Conclusion}
\label{sec:conclusion}
In this paper, we conducted a comprehensive study of RL algorithms applied to the gym-DSSAT crop simulator environment. We provided an overview of the PPO and DQN algorithms, detailed the gym-DSSAT environment, and performed numerical experiments to evaluate the ability of PPO and DQN to learn fertilization and irrigation policies. Our results showed that PPO outperformed DQN in individual tasks, such as fertilization and irrigation, but struggled to outperform DQN when tasked with managing both simultaneously. As discussed, one limitation of our study was the lack of parameter tuning during training, which likely impacted performance. Future work should focus on optimizing hyperparameters for both PPO and DQN, as well as exploring alternative reward functions and crop management scenarios. We also suggest that future research consider applying offline RL algorithms for crop management. Offline RL, which relies on static datasets rather than real-time interactions, aligns more closely with how farmers operate and has the potential to deliver more practical solutions in real-world farming applications \citep{levine2020offline}.

\section*{Authorship Contribution}
\textbf{Joseph Balderas}: Investigation, Software, Formal analysis, Writing - original draft;
\textbf{Dong Chen}: Conceptualization, Methodology, Resources, Supervision, Writing - review \& editing; 
\textbf{Yanbo Huang}: Conceptualization, Resources, Supervision, Writing - review \& editing;
\textbf{Li Wang}: Resources, Supervision, Writing - review \& editing;
\textbf{Ren-Cang Li}: Resources, Supervision, Writing - review \& editing.

\section*{Acknowledgement}
This research was financially supported by a summer intern of a Ph.D. student at the University of Texas at Arlington through the USDA ARS Research Apprenticeship Program at the University of Texas at Arlington through a Non- Assistance Cooperative Agreement (Project Number: 6066-21310-006-021-S). The USDA-ARS scientist works under the federal in-house appropriated project (Project Number: 6064-21600-001-000-D) from USDA-ARS National Program 216 - Sustainable Agricultural Systems. Additionally, the project was supported in part by NSF DMS-2407692, NIH R01AG075582, and NIH R21AG079309.

\typeout{}
\bibliography{ref}
\end{document}